\documentclass[onecollarge]{svjour2}
\usepackage{graphicx}
\usepackage{cite}
\usepackage{color}
\usepackage{amsfonts}
\usepackage{soul}
\usepackage{cancel}
\usepackage{amsmath}

\begin{document}

\title{Rate of mutual information between coarse-grained non-Markovian variables}

\author{Andre C Barato \and David Hartich \and Udo Seifert}

\institute{
A. C. Barato 
\at II. Institut f\"ur Theoretische Physik, Universit\"at Stuttgart - Stuttgart 70550, Germany\\
\email{barato@theo2.physik.uni-stuttgart.de}\\
\and
D. Hartich 
\at  II. Institut f\"ur Theoretische Physik, Universit\"at Stuttgart - Stuttgart 70550, Germany\\
\email{hartich@theo2.physik.uni-stuttgart.de}\\
\and
U. Seifert
\at II. Institut f\"ur Theoretische Physik, Universit\"at Stuttgart - Stuttgart 70550, Germany\\
\email{useifert@theo2.physik.uni-stuttgart.de}\\
}

\date{Received: date / Accepted: date}

\maketitle
\begin{abstract}
The problem of calculating the rate of mutual information between two coarse-grained variables that together specify a continuous time Markov process is addressed.
As a main obstacle, the coarse-grained variables are in general non-Markovian, therefore, an expression for their Shannon entropy rates
in terms of the stationary probability distribution is not known. A numerical method to estimate the Shannon entropy rate of continuous time hidden-Markov processes from a single time series is developed.
With this method the rate of mutual information can be determined numerically. Moreover, an analytical upper bound on the rate of mutual information is calculated for a class of Markov processes for which the transition
rates have a bipartite character. Our general results are illustrated with explicit calculations for four-state networks.

 
\end{abstract}

\def\conf{c}					
\def\rate{w}					
\def\headline#1{\paragraph{\bf #1\\[2mm]}}	
\def\multiindex#1{\mathbf{#1}}			
\def\d{{\rm d}}
\def\Ps{{P_{\scriptscriptstyle \hspace{-0.3mm} s}}}
\def\MF{{\mbox{\tiny \rm \hspace{-0.3mm} MF}}}
\def\i{\mathcal{I}} 

\newpage
\parskip 2mm 
\pagestyle{plain}

\section{Introduction}

Mutual information \cite{shan48,cove06} is a quantity of central importance in information theory. It is a nonlinear correlation function \cite{li90} between two random variables 
that measures how much information about one random variable is encoded in the other. In other words, it measures the reduction of the uncertainty of a random variable resulting from knowing the other one.
Since the Shannon entropy quantifies the randomness of a random variable, mutual information is a difference between Shannon entropies. More generally, given two stochastic time series 
the information per unit of time between them is quantified by the rate of mutual information, which is a difference between Shannon entropy rates. Whereas the Shannon entropy rate of Markovian time
series can be expressed in terms of the stationary probability distribution \cite{cove06}, no general formula is known for non-Markovian processes.

Recently, we have obtained an analytical upper bound on the rate of mutual information and calculated it numerically for a class of Markov processes \cite{bara13}. This class is formed by bipartite networks
where the full state of the systems is determined by two coarse-grained variables: one corresponding to an external Markovian process and the other to an internal non-Markovian process. In this paper we generalize the results obtained in \cite{bara13}
by calculating an upper bound on the rate of mutual information for a more general class of Markov processes, where both coarse-grained processes can be non-Markovian. Moreover, we develop a numerical method to estimate the Shannon entropy rate 
of a continuous time coarse-grained non-Markovian process by adapting an extant numerical method for discrete time \cite{holl06,jacq08,rold12}.

Apart from the quite challenging mathematical problem of determining the rate of mutual information, there are physical motivations for our study. First, within
stochastic thermodynamics \cite{seif12}, which is a framework for far from equilibrium systems, a central quantity is the thermodynamic entropy production. In a nonequilibrium steady state, it characterizes 
the rate at which heat is dissipated. On the other hand, the rate of mutual information is an information theoretic entropy rate that characterizes the correlations between the two coarse-grained processes.
The study of the relation between both quantities for specific models should improve our understanding of the relation between thermodynamics and information.

More specifically, a considerable amount of work on the role of information in the stochastic thermodynamics of feedback driven systems, for which a controller acts at periodic time intervals, has emerged recently 
\cite{touc00,cao09,saga10,saga12b,saga12,horo10,horo11,gran11,abre11a,abre11,baue12,ito11,cris12,mand12}. In such periodic steady states the rate of mutual information between system and controller 
is just the average mutual information due to each new measurement divided by the length of the period \cite{saga12}. The second law of thermodynamics bounding the maximum extractable work has then to be modified in order to include the 
mutual information between system and controller, linking directly thermodynamic and information theoretic entropy productions. On the other hand, 
if a Maxwell's demon is described as an autonomous system \cite{espo12b,stra13,bara13a}, calculating the rate of mutual information in such a genuine nonequilibrium steady states shows that in this case there is
no such relation between the rate of mutual information and the thermodynamic entropy production \cite{bara13}.  

Second, the study of the energetic costs of sensing in biochemical networks is a field emerging at this interface between 
thermodynamics and information theory\cite{lan12,meht12}. For example, an intriguing 
relation between the energy costs of dissipation, quantified by the thermodynamic entropy production, and the adaptation error has been found in a model for the {\sl E. coli} sensory system \cite{lan12}. In these papers, the observables characterizing  
the quality of sensing are the adaptation error \cite{lan12} and the uncertainty in the external ligand concentration \cite{meht12}. Alternatively, a natural quantity that should be discussed in this context with the same 
dimension of the thermodynamic entropy production is
the rate of mutual information. Hence, the study of the relation between these two quantities in biochemical sensory networks could contribute
to an understanding of the thermodynamics of such systems \cite{bara13}.

This paper is organized as follows. In Sect. \ref{sec2}, we discuss a one spin system with a fluctuating magnetic field as a simple introductory example. We define the bipartite network and the quantities of interest in Sect. \ref{sec3}.
In Sect. \ref{sec4}, we derive our first main result, which is the analytical upper bound on the rate of mutual information. Our second main result, namely, the continuous time numerical method, is explained in
Sect. \ref{sec5}, where we also discuss the discrete time case. In Sect. \ref{sec6}, we calculate the rate of mutual information explicitly for four-state systems considering cases  where the rate of mutual information
admits a simple interpretation. We conclude in Sect. \ref{sec7}.

\section{One spin out of equilibrium}
\label{sec2}

\begin{figure}[h]
\centering
\includegraphics{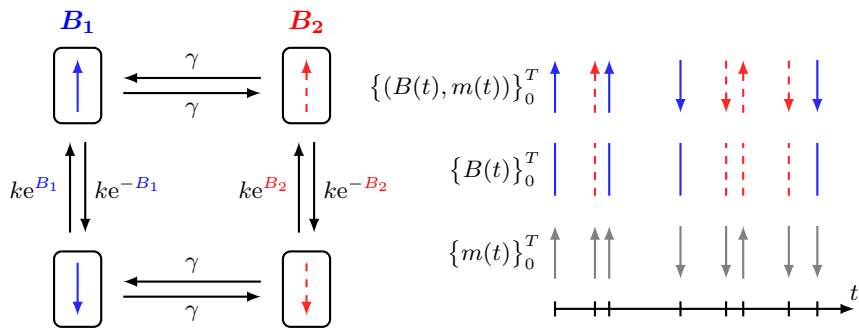}
\caption{One spin system in a time varying magnetic field. The transition rules for the model are shown in the left panel. The vertical transitions correspond to a spin flip due to thermal fluctuations and fulfill local detailed balance. The horizontal
transitions at rate $\gamma$, controlled by an external device, correspond to a change in the magnetic field between the values $B_1$ (full blue line) and $B_2$ (dashed red line). The right panel shows the corresponding time series.
}
\label{fig1}
\end{figure}

For a simple illustration let us start with the four-state model represented in Fig. \ref{fig1}. One spin is subjected to a time varying magnetic field while in contact with a thermal reservoir inducing 
flips. The magnetic field is controlled by some external device that randomly changes it. More precisely, the field is a Poisson process with rate $\gamma$, fluctuating between the values $B_1$ and $B_2$.
The transition rates for the spin flip are denoted by $w^{\alpha}_{mm'}$ (from $m$ to $m'$), where $\alpha=1,2$ represents the state of the magnetic field and $m,m'=-,+$ the orientation of the spin.
These transition rates are given by the local detailed balance assumption, i.e., 
\begin{equation}
\frac{w^{\alpha}_{+-}}{w^{\alpha}_{-+}}=\exp(-2 B_\alpha),
\end{equation}
where we set Boltzmann constant multiplied by temperature to $1$.

Now consider the three time series shown in Fig. \ref{fig1}. The first time series $\{(B(t),m(t))\}_0^T$ represents a stochastic trajectory of the full four-state Markov process. The time series $\{B(t)\}_0^T$ is also Markovian, because 
if we integrate out the spin variable we get a two-state Markov process. The physical reason for the Markov character of this process is that the magnetic field is controlled by an external device that does not care about the 
internal state (the spin orientation). The spin time series $\{m(t)\}_0^T$ is non-Markovian and contains information about the magnetic field time series $\{B(t)\}_0^T$, i.e., both are correlated.

The rate of mutual information $\i$ (see definition below) quantifies how much information about the time series $\{B(t)\}_0^T$ is encoded in the time series $\{m(t)\}_0^T$. In other words, it gives 
a (non-linear) measure of how correlated both time series are, being zero in the case where they are independent and positive otherwise. Within the present model, both time series become independent only 
for $B_1=B_2$. For this choice of parameters, we obtain a two-state Markov process for the spin by integrating out the magnetic field, i.e., for $B_1=B_2$ the processes  $\{B(t)\}_0^T$ and $\{m(t)\}_0^T$ become two independent Markov processes.

Moreover, the model is in equilibrium, i.e., detailed balance is fulfilled if and only if $B_1=B_2$. The thermodynamic entropy production $\sigma$ (see definition below) 
is a signature of nonequilibrium since it is zero when detailed balance is fulfilled and 
strictly positive for nonequilibrium stationary states. For this one-spin system, $\sigma$ is the rate at which the system dissipates heat to the thermal reservoir.

\begin{figure}
\centering
\includegraphics{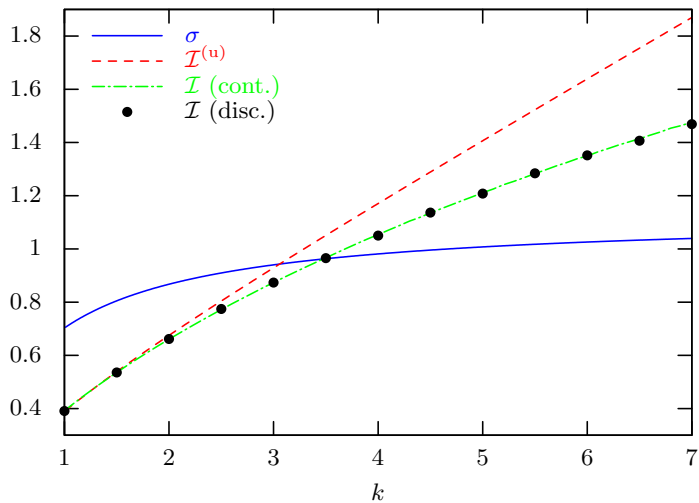}
\caption{Comparison of the thermodynamic entropy production $\sigma$ (\ref{sdef}) and the rate of mutual information $\i$ for the model of Fig. \ref{fig1} as a function of $k$, where $\gamma=1$, $B_1=0$ and $B_2=\ln(10)$. The abbreviation (disc.) 
indicates the mutual information obtained with the extrapolation for $\tau\to0$ in discrete time explained in Sect. \ref{5.1} and (cont.) is related to the continuous time numerical method explained Sect. \ref{5.2}. $\i^{(u)}$
shows the analytical upper bound (\ref{ubi}).} 
\label{fig3}
\end{figure}

As cited in the introduction, for feedback driven systems the second law of thermodynamics has to be adapted in order to include the rate of mutual information between the system and the controller.
For these systems it is possible to rectify fluctuations in order to extract work from a single heat bath, where the rate of the extracted work is bounded by the rate of mutual information. A complementary question, considering
the model of Fig. \ref{fig1}, is whether the rate of mutual information between $m(t)$ and $B(t)$, which is non-zero only when the system is out of equilibrium, is bounded by the  dissipation rate required  
to sustain the nonequilibrium stationary state. In \cite{bara13} we have shown that, in general, there is no such bound. In Fig. \ref{fig3}, we compare the thermodynamic entropy production $\sigma$ with 
the rate of mutual information $\i$ for the one-spin system of Fig. \ref{fig1}  using the results derived further below.

\section{Bipartite network}
\label{sec3}
We now define the class of bipartite Markov processes studied in this paper and the rate of mutual information, for discrete and continuous time.

\subsection{Shannon entropy rates for discrete time}

First we consider a discrete time Markov process where the states are labeled by the pair of variables $(\alpha,i)$, where $\alpha=1,\ldots,\Omega_x$ and $i=1,\ldots,\Omega_y$. 
The Markov chain is defined by the following transition probabilities,   
\begin{equation}	
W_{ij}^{\alpha\beta}\equiv\left\{
\begin{array}{ll} 
 w^{\alpha\beta}_i\tau  \quad &\textrm{if $i=j$ and $\alpha\neq\beta$}, \\
 w^{\alpha}_{ij}\tau  \quad &\textrm{if $i\neq j$ and $\alpha=\beta$},\\
 0   \quad &\textrm{if $i\neq j$ and $\alpha\neq\beta$},\\
 1-\sum_{k\neq i} w^{\alpha}_{ik}\tau-\sum_{\gamma\neq\alpha}w^{\alpha\gamma}_{i}\tau  \qquad  &\textrm{if $i=j$ and $\alpha=\beta$}, 
\end{array}\right.\,
\label{defrates}
\end{equation}
where $\tau$ is the time spacing. Transitions where both variables change are not allowed, which means that the network of states is bipartite.

We denote a discrete time series of the full Markov process with $N$ jumps by $\{Z_n\}_0^N= (Z_0,Z_1,\ldots,Z_N)$, where $Z_n= (X_n,Y_n)\in \{1,\ldots,\Omega_x\}\times\{1,\ldots,\Omega_y\}$.
The Shannon entropy rate of the Markov chain (\ref{defrates}) is defined by \cite{cove06}  
\begin{equation}
H_{Z}\equiv -\lim_{N\to \infty}\frac{1}{N\tau}\sum_{\{Z_n\}_0^N}P[\{Z_n\}_0^N]\ln P[\{Z_n\}_0^N],
\label{entZ}
\end{equation}
where the sum is over all possible stochastic trajectories $\{Z_n\}_0^N$. 
Since the full process is Markovian, it is well known that  this entropy rate can be expressed in terms of the stationary probability distribution $P_i^{\alpha}$ in the form \cite{cove06}
\begin{equation}
H_{Z}= -\frac{1}{\tau}\sum_{i,j,\alpha,\beta}P^\alpha_i W_{ij}^{\alpha\beta}\ln W_{ij}^{\alpha\beta}.
\label{entZ2}
\end{equation}

Moreover, the Shannon entropy rates of the coarse-grained processes $\{X_n\}_0^N$ and $\{Y_n\}_0^N$ are defined as  
\begin{equation}
H_{X}\equiv -\lim_{N\to \infty}\frac{1}{N\tau}\sum_{\{X_n\}_0^N}P[\{X_n\}_0^N]\ln P[\{X_n\}_0^N],
\label{entX}
\end{equation}
\begin{equation}
H_{Y}\equiv -\lim_{N\to \infty}\frac{1}{N\tau}\sum_{\{Y_n\}_0^N}P[\{Y_n\}_0^N]\ln P[\{Y_n\}_0^N].
\label{entY}
\end{equation}
These two coarse-grained processes are in general non-Markovian. More precisely, they are hidden Markov processes \cite{ephr02}. The quantity we wish to calculate is the rate of mutual information between the two
coarse-grained variables, which is defined as 
\begin{equation}
I \equiv H_X+H_Y-H_{Z}.
\label{mutualdef}
\end{equation}
Therefore, in order to obtain the rate of mutual information we have to calculate the Shannon entropy rates of the two coarse-grained variables $X$ and $Y$. Using the definitions of the Shannon entropy rates
(\ref{entZ}), (\ref{entX}), and (\ref{entY}), we can rewrite $I $ in the form 
\begin{equation}
I = \lim_{N\to \infty}\frac{1}{N\tau} D_{KL}(P[\{Z_n\}_0^N]||P[\{X_n\}_0^N]P[\{Y_n\}_0^N]),
\end{equation}
where the Kullback-Leibler distance is defined as \cite{cove06}
\begin{equation}
D_{KL}(P[\{Z_n\}_0^N]||P[\{X_n\}_0^N]P[\{Y_n\}_0^N])\equiv\sum_{\{Z_n\}_0^N}P[\{Z_n\}_0^N]\ln\frac{P[\{Z_n\}_0^N]}{P[\{X_n\}_0^N]P[\{Y_n\}_0^N]}.
\end{equation}
With this formula it becomes explicit that the rate of mutual information measures how correlated the two processes are. 

In \cite{bara13} we have studied the particular case where $X$ is an external process independent of the internal states, i.e., $w^{\alpha\beta}_i\equiv w^{\alpha\beta}$ for all $i=1,\ldots,\Omega_y$. In this case, the external process is
also Markovian and $H_{X}$ becomes
\begin{equation}
H_X= -\frac{1}{\tau}\sum_{\alpha,\beta}P^\alpha W^{\alpha\beta}\ln W^{\alpha\beta},
\label{entx2}
\end{equation}
where 
\begin{equation}
P^\alpha\equiv \sum_{i=1}^{\Omega_y} P_i^\alpha.
\end{equation}
For later convenience we also define  
\begin{equation}
P_i\equiv \sum_{\alpha=1}^{\Omega_x} P_i^\alpha.
\end{equation}

In this paper we are mostly interested in the continuous time limit $\tau\to0$. For the calculation of an analytical upper bound on the continuous time rate of mutual information, it is useful to consider the discrete time Markov chain and then take the 
limit $\tau\to 0$. In this limit, the Shannon entropy rates diverge as $\ln \tau$ \cite{gasp04a,leco07a}, however, the rate of mutual information is a well defined finite quantity: it is a difference between Shannon entropy rates 
for which the term proportional to $\ln \tau$ cancels. 

It is possible to calculate the rate of mutual information numerically for the discrete time case as a function of $\tau$ and then extrapolate to the limit $\tau\to0$. Alternatively, we develop a more efficient numerical method to directly 
estimate the entropy rate of a continuous time series. We now define the Shannon entropy rates and the rate of mutual information for the continuous time case.   

\subsection{Shannon entropy rates for continuous time}

The continuous time Markov process is defined by the transition rates (transition probability per time)
\begin{equation}	
w_{ij}^{\alpha\beta}\equiv\left\{
\begin{array}{ll} 
 w^{\alpha\beta}_i & \quad \textrm{if $i=j$ and $\alpha\neq\beta$}, \\
 w^{\alpha}_{ij} & \quad  \textrm{if $i\neq j$ and $\alpha=\beta$},\\
 0 & \quad \textrm{if $i\neq j$ and $\alpha\neq\beta$}. 
\end{array}\right.\,
\label{defrates2}
\end{equation}
The stochastic trajectory for a fixed time interval $T$ is written as $\{Z(t)\}_0^T$ (in this case the time interval is fixed and the number of jumps $N$ is a random variable). 
Similarly, the definition of the Shannon entropy rate of the full Markov process is   
\begin{equation}
\mathcal{H}_{Z}\equiv -\lim_{T\to \infty}\frac{1}{T}\int D[\{Z(t)\}_0^T]\mathcal{P}[\{Z(t)\}_0^T]\ln\mathcal{P}[\{Z(t)\}_0^T],
\label{entropyzcont}
\end{equation}
where $\mathcal{P}[\{Z(t)\}_0^T]$ is the probability density of the trajectory $\{Z(t)\}_0^T$ and the integral is over all possible stochastic trajectories. Since the $Z$ process is Markovian the continuous time Shannon entropy 
rate can also be written in terms of the stationary probability distribution, and it is given by \cite{dumi88}  
\begin{equation}
\mathcal{H}_{Z}=  -\sum_{i,\alpha}P^\alpha_i\sum_{j,\beta \neq i,\alpha }  w_{ij}^{\alpha\beta}(\ln w_{ij}^{\alpha\beta}-1).
\label{entropyzcontform}
\end{equation}
Since the transition rates can take any positive value, it is clear that this Shannon entropy rate can be negative. This is a well known fact for continuous random variables \cite{cove06}. 
The Shannon entropy rates of the $X$ and $Y$ processes are defined in the same way,
\begin{equation}
\mathcal{H}_{X}\equiv -\lim_{T\to \infty}\frac{1}{T}\int D[\{X(t)\}_0^T]\mathcal{P}[\{X(t)\}_0^T]\ln\mathcal{P}[\{X(t)\}_0^T],
\label{entropyxcont}
\end{equation}
\begin{equation}
\mathcal{H}_{Y}\equiv -\lim_{T\to \infty}\frac{1}{T}\int D[\{Y(t)\}_0^T]\mathcal{P}[\{Y(t)\}_0^T]\ln\mathcal{P}[\{Y(t)\}_0^T].
\label{entropyycont}
\end{equation}
Moreover, the definition of the continuous time rate of mutual information is 
\begin{equation}
\i\equiv \mathcal{H}_X+\mathcal{H}_Y-\mathcal{H}_{Z},
\label{mutualdef2}
\end{equation}
where the relation between $\i$ and the discrete time rate of mutual information (\ref{mutualdef}) is $\i= \lim_{\tau\to 0}I $. 
Even though the Shannon entropy rates $\mathcal{H}_X$, $\mathcal{H}_Y$, and $\mathcal{H}_Z$  may be negative, the rate of mutual information $\i$,  the quantity of central interest in this paper, 
fulfills $\i\ge0$. In order to show this we write the rate of mutual information as a Kullback-Leibler distance, 
\begin{equation}
\i=  \lim_{T\to \infty}\frac{1}{T} D_{KL}\left(\mathcal{P}[\{Z(t)\}_0^T]||\mathcal{P}[\{X(t)\}_0^T]\mathcal{P}[\{Y(t)\}_0^T]\right)\ge 0.
\end{equation}

\subsection{Thermodynamic entropy production}

A central quantity in stochastic thermodynamics is the thermodynamic entropy production \cite{seif12,schn76}, which for the rates (\ref{defrates2}) reads
\begin{equation}
\sigma\equiv \sum_{i,\alpha}P_i^\alpha\left(\sum_{j\neq i} w^\alpha_{ij}\ln\frac{w^\alpha_{ij}}{w^\alpha_{ji}}+\sum_{\beta\neq\alpha} w^{\alpha\beta}_i\ln \frac{w^{\alpha\beta}_i}{w^{\beta\alpha}_i}\right).
\label{sdef}
\end{equation} 
Analogously to the rate of mutual information, the thermodynamic entropy production can also be expressed as \cite{kawa07} 
\begin{equation}
\sigma\equiv  \lim_{T\to \infty}\frac{1}{T} D_{KL}\left(\mathcal{P}[\{Z(t)\}_0^T]||\mathcal{P}[\{\tilde{Z}(t)\}_0^T]\right)\ge 0.
\end{equation}
where $\{\tilde{Z}(t)\}_0^T$ denotes the time-reversed trajectory, i.e., $\tilde{Z}(t)= Z(T-t)$. Depending on the physical interpretation of the transition rates, the entropy rate $\sigma$ may characterize the dissipation associated 
with the full network of states, being zero only if detailed balance is fulfilled. As discussed above, for the  
one spin system of Fig. \ref{fig1} it is proportional to the heat that flows from the system to the thermal reservoir. On the other hand,
$\i$ is the information theoretic entropy rate that quantifies the correlation between the $X$ and $Y$ processes. No closed formula like equation (\ref{sdef}) is known for the rate of mutual information. However, as we show next,
it is still possible to calculate it numerically and to obtain an analytical upper bound.

\section{Analytical upper bound}
\label{sec4}

Let us take the $Y$ process in the discrete time case and in the stationary regime. The conditional Shannon entropy is defined as
\begin{equation}
H(Y_{N+1}|Y_{N},\ldots,Y_1)\equiv \frac{1}{\tau}\sum_{Y_N+1,Y_N,\ldots,Y_1} P(Y_{N+1},Y_N,\ldots,Y_1)\ln P(Y_{N+1}|Y_N,\ldots,Y_1),
\label{conddef}
\end{equation}
where $P(Y_{N+1}|Y_N,\ldots,Y_1)= P(Y_{N+1},Y_N,\ldots,Y_1)/P(Y_N,\ldots,Y_1) $ is a conditional probability. The knowledge of one extra random variable can only decrease the uncertainty about $Y_{N+1}$, which means that
$H(Y_{N+1}|Y_{N},\ldots,Y_2,Y_1)\le H(Y_{N+1}|Y_{N},\ldots,Y_2)$. Therefore, as the $Y$ process is stationary, we obtain that this conditional entropy is a decreasing function of $N$, i.e., 
\begin{equation}
H(Y_{N+1}|Y_{N},\ldots,Y_1)\le H(Y_{N}|Y_{N-1},\ldots,Y_1).
\label{conddec}
\end{equation}
Moreover, in the limit $N\to\infty$, we have \cite{cove06} 
\begin{equation} 
\lim_{N\to\infty} H(Y_{N+1}|Y_{N},\ldots,Y_1)= H_Y,
\end{equation} 
which means that the conditional entropy (\ref{conddef}) bounds the Shannon entropy rate $H_Y$ from above. Furthermore, it can be shown that $H_Y$ is bounded from below by \cite{cove06}
\begin{equation}
H(Y_{N+1}|Y_{N},\ldots,Y_2,Z_1)=H(Y_{N+1}|Y_{N},\ldots,Y_2,Y_1,X_1),
\end{equation}
leading to
\begin{equation}
H(Y_{N+1}|Y_{N},\ldots,Y_1,X_1)\le H_Y\le H(Y_{N+1}|Y_{N},\ldots,Y_1),
\label{bounds}
\end{equation}
where the bounds become tighter for increasing $N$.

As we show in the appendix, for any finite $N$, 
\begin{eqnarray}
H(Y_{N+1}|Y_{N},\ldots,Y_1)=-\sum_{i,\alpha}P_i^\alpha\sum_{j\neq i}w_{ij}^\alpha\left(\ln \tau+\ln\frac{\sum_\beta P_i^\beta w_{ij}^\beta}{P_i}-1\right)+\textrm{O}(\tau),
\label{hyN}
\end{eqnarray} 
and, analogously, 
\begin{eqnarray}
H(X_{N+1}|X_{N},\ldots,X_1)=-\sum_{i,\alpha}P_i^\alpha\sum_{\beta\neq \alpha}w_{i}^{\alpha\beta}\left(\ln \tau+\ln\frac{\sum_j P_j^\alpha w_{j}^{\alpha\beta}}{P^\alpha}-1\right)+\textrm{O}(\tau).
\label{hxN}
\end{eqnarray} 
From (\ref{entZ2}) we obtain the following formula for the entropy rate $H_Z$, 
\begin{equation}
H_Z= -\sum_{i,\alpha}P_i^\alpha\left(\sum_{j\neq i}w^\alpha_{ij}(\ln \tau+\ln w^\alpha_{ij}-1)+\sum_{\beta\neq \alpha}w^{\alpha\beta}_{i}(\ln \tau+\ln w^{\alpha\beta}_{i}-1)\right)+\textrm{O}(\tau).
\label{hzN}
\end{equation}
For convenience we define the average transition rates
\begin{equation}
\overline{w_{ij}}\equiv \sum_{\alpha=1}^{\Omega_x}P(\alpha|i) w_{ij}^{\alpha}= \frac{1}{P_i}\sum_{\alpha=1}^{\Omega_x}P_i^\alpha w_{ij}^{\alpha}, 
\label{avgi}
\end{equation}
\begin{equation}
\overline{w^{\alpha\beta}}\equiv \sum_{i=1}^{\Omega_y}P(i|\alpha) w_{i}^{\alpha\beta}=  \frac{1}{P^\alpha}\sum_{i=1}^{\Omega_y}P_i^\alpha w_{i}^{\alpha\beta}. 
\label{avgalpha}
\end{equation}

The $N$-th upper bound on the rate of mutual information is then 
\begin{equation}
I ^{(u,N)}  \equiv H(Y_{N+1}|Y_{N},\ldots,Y_1)+H(X_{N+1}|X_{N},\ldots,X_1)-H_Z.
\end{equation}
From equations (\ref{hyN}), (\ref{hxN}), and (\ref{hzN}), it is given by
\begin{align}
I ^{(u,N)} = \sum_{i,\alpha}P_i^\alpha\left(\sum_{j\neq i}w^\alpha_{ij}\ln \frac{w^\alpha_{ij}}{\overline{w_{ij}}}+\sum_{\beta\neq \alpha}w^{\alpha\beta}_{i}\ln 
\frac{w^{\alpha\beta}_{i}}{\overline{w^{\alpha\beta}}}\right)+\textrm{O}(\tau).
\end{align}
Taking the continuous time limit $\tau\to 0$, the rate of mutual information is hence bounded from above by
\begin{align}
\i^{(u)}\equiv \sum_{i,\alpha}P_i^\alpha\left(\sum_{j\neq i}w^\alpha_{ij}\ln \frac{w^\alpha_{ij}}{\overline{w_{ij}}}+\sum_{\beta\neq \alpha}w^{\alpha\beta}_{i}\ln 
\frac{w^{\alpha\beta}_{i}}{\overline{w^{\alpha\beta}}}\right).
\label{ubi}
\end{align}
Two remarks are important. First, it is interesting to note the formal similarity between this expression and the one for the thermodynamic entropy production (\ref{sdef}). Substituting in the latter inside the
logarithm the rate of a reversed transition by the respective average forwards rates (\ref{avgi}) and (\ref{avgalpha}), we get the former. Second, to calculate the true rate of mutual information we would have to take the 
limit $N\to \infty$ with fixed $\tau$. This would give an expression for the rate of mutual information that would be valid for any time spacing $\tau$ and should become the continuous time rate of mutual information
by taking the limit $\tau\to 0$ afterwards.

\begin{figure}
\centering
\includegraphics{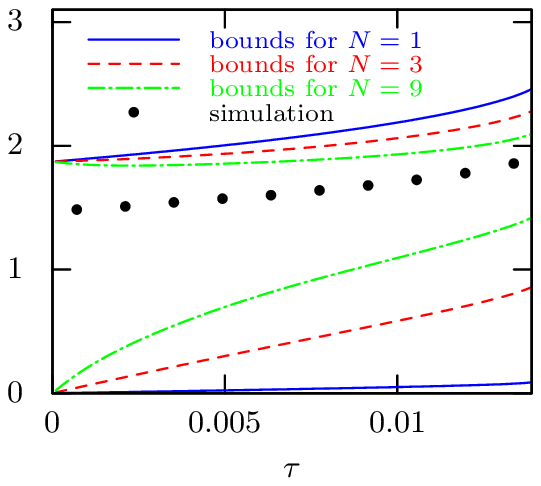}
\includegraphics{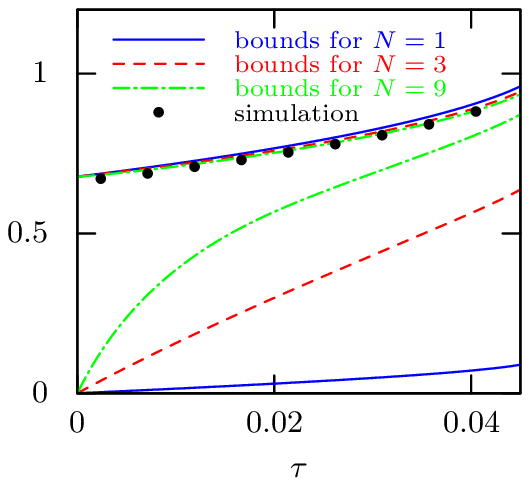}
\caption{Lower and upper bounds for $N=1,3,9$ obtained from (\ref{bounds}) for the discrete time version of the model of Fig. \ref{fig1}. The parameters are $\gamma=1$, $B_1=0$, $B_2=\ln(10)$,
$k=7$ (left panel), and $k=2$ (right panel). In the limit $\tau\to 0$, the upper bounds go to the value given by (\ref{ubi}) and the lower bounds go to zero.} 
\label{fig4}
\end{figure}

A similar calculation for the lower bounds in equation (\ref{bounds}) shows that in the continuous time limit they all go to zero. This is illustrated in Fig. \ref{fig4}, where
we plot upper and lower bounds obtained from (\ref{bounds}) as a function of the time spacing $\tau$ for the discrete time version of the one spin model of Fig. \ref{fig1}. This discrete time version is defined by the transition probabilities 
given by (\ref{defrates}) obtained from the transition rates represented in Fig. \ref{fig1}.

Finally, one limiting case for which the rate of mutual information saturates the upper bound is the following. We take the $X$ process to be Markovian, i.e., $w^{\alpha\beta}_i\equiv w^{\alpha\beta}$ for all $i=1,\ldots,\Omega_y$. From
equation (\ref{entx2}), it follows
\begin{equation}
H_X=  -\sum_{\alpha}P^\alpha\left(\sum_{\beta\neq \alpha}w^{\alpha\beta}(\ln \tau+\ln w^{\alpha\beta}-1)\right)+\textrm{O}(\tau).
\label{hxmark}
\end{equation}
Furthermore, if the $X$ transitions are much faster than the $Y$ transitions ($w^{\alpha\beta}\gg w^{\alpha}_{ij}$), the $Y$ process becomes approximately Markovian, with transition rates $\overline{w_{ij}}$ \cite{raha07,espo12}.  
Therefore, in this limit we expect 
\begin{equation}
H_Y=  -\sum_{i}P_i\left(\sum_{j\neq i}\overline{w_{ij}}(\ln \tau+\ln \overline{w_{ij}}-1)\right)+\textrm{O}(\tau).
\label{hymark}
\end{equation}   
The continuous time rate of mutual information $\i$ obtained from (\ref{hzN}), (\ref{hxmark}) and (\ref{hymark}) is then precisely the upper bound (\ref{ubi}). Therefore, in the case where the $X$ process is Markovian and much faster then
the $Y$ process, the rate of mutual information saturates the upper bound (\ref{ubi}). In Sect. \ref{sec6}, we illustrate this fact explicitly for four-state models.

\section{Estimating Shannon entropy rate from a single time series}
\label{sec5}
\subsection{Discrete time}
\label{5.1}
For discrete time, the probability of a stochastic trajectory of the $Y$ process can be written as 
\begin{equation}
P[\{Y_n\}_{0}^{N}]= \sum_{X_{N}X_{N-1}\ldots X_1 X_0} P[Y_N,X_N|Y_{N-1},X_{N-1}] \ldots P[Y_1,X_1|Y_{0},X_{0}] P(X_0,Y_0),
\label{masternon}
\end{equation}   
where $P(X_0,Y_0)$ denotes the initial probability distribution and $P[X_n,Y_n|X_{n-1},Y_{n-1}]$ is the conditional probability. Explicitly, 
for $(X_{n-1},Y_{n-1})=(\alpha,i)$ and $(X_{n},Y_{n})=(\beta,j)$ we have $P[X_n,Y_n|X_{n-1},Y_{n-1}]=W^{\alpha\beta}_{ij}$. 

Let the random matrix $\boldsymbol{T}(Y_n,Y_{n-1})$ be defined by 
\begin{equation}
\boldsymbol{T}(Y_n,Y_{n-1})_{X_n,X_{n-1}}\equiv P[X_n,Y_n|X_{n-1},Y_{n-1}]=P[Z_n|Z_{n-1}].
\label{defTy}
\end{equation}
This is a $\Omega_x\times\Omega_x$ matrix, where the variables $(Y_n,Y_{n-1})$ make it random. Using this matrix, equation (\ref{masternon}) can be rewritten as 
\begin{equation}
P[\{Y_{n}\}_{n=0}^{N\tau}]=  \mathbf{V} \boldsymbol{T}(Y_N,Y_{N-1})\ldots \boldsymbol{T}(Y_1,Y_{0}) \mathbf{P}_{Y_0}
\label{prodY}
\end{equation}
where $\mathbf{V}$ is a row vector with all $\Omega_x$ components equal to one and $\mathbf{P}_{Y_0}$ is a column vector with components $P(Y_0,X_0)$, with $X_0= 1,\ldots,\Omega_x$. The Shannon entropy rate (\ref{entY})
can then be written as 
\begin{equation}
H_Y= -\lim_{N\to\infty} \frac{1}{N\tau}\sum_{Y_N,Y_{N-1},\ldots,Y_0} P(Y_{N},Y_{N-1},\ldots,Y_0)\ln \mathbf{V} \boldsymbol{T}(Y_N,Y_{N-1})\ldots \boldsymbol{T}(Y_1,Y_{0}) \mathbf{P}_{Y_0}.
\label{prodY2}
\end{equation}
Moreover, in the large $N$ limit, where boundary terms become irrelevant, we can replace the product of matrices (\ref{prodY}) in equation (\ref{prodY2}) with $\left\lVert \prod_{n=1}^{N}\boldsymbol{T}(Y_n,Y_{n-1})\right\rVert$, 
where $ \lVert\cdot \rVert$ is any matrix norm \cite{holl06}. Therefore, in order to estimate the entropy rate $H_Y$ we generate a long time series $\{Y^{*}_n\}_{0}^{N}$ with a numerical simulation and calculate 
\begin{equation}
H_Y\simeq -\frac{1}{N\tau} \ln \left\lVert \prod_{n=1}^{N}\boldsymbol{T}(Y^{*}_n,Y^{*}_{n-1})\right\rVert.
\label{eq41}
\end{equation}
Such a numerical method to calculate the Shannon entropy rate has been used in \cite{jacq08,holl06,rold12}. The appropriate way to calculate this product, avoiding numerical precision problems
for large $N$, is to normalize the product every $L$ steps and repeat the procedure $M$ times, so that $N=ML$ \cite{cris93}. More precisely, for $m=1,\ldots,M$ we calculate the vector
\begin{equation}
\mathbf{v}_m= \left[\prod_{l=(m-1)L+1}^{mL} \boldsymbol{T}(Y_l^*,Y_{l-1}^*)\right] \mathbf{u}_{m-1},
\end{equation} 
and the normalization factor
\begin{equation}
R_m= \left\lVert \mathbf{v}_m \right \rVert,
\end{equation} 
where $\mathbf{u}_{m}$ is the normalized vector 
\begin{equation}
\mathbf{u}_m= \frac{\mathbf{v}_m}{R_m},
\end{equation} 
and the initial vector $\mathbf{u}_0$ is any random vector with an unitary norm. By calculating the normalization factors iteratively we obtain the Shannon entropy rate with the formula
\begin{equation}
H_Y\simeq -\frac{1}{ML\tau} \sum_{m=1}^{M}\ln  R_m.
\end{equation}

The present method is based on the fact that the probability of an $Y$ stochastic trajectory can be written as a product of random matrices (\ref{prodY}). Since this is true for any coarse-grained non-Markovian variable we can also apply 
the same method to calculate $H_X$. Explicitly, if we define the  $\Omega_y\times\Omega_y$ random matrix
\begin{equation}
\boldsymbol{T}(X_n,X_{n-1})_{Y_n,Y_{n-1}}\equiv P[X_n,Y_n|X_{n-1},Y_{n-1}],
\label{defTx}
\end{equation}
then we can estimate the Shannon entropy rate from the numerically generated time series $\{X^*_n\}_{0}^{N}$ from 
\begin{equation}
H_X\simeq -\frac{1}{N\tau} \ln \left\lVert \prod_{n=1}^{N}\boldsymbol{T}(X^*_n,X^*_{n-1})\right\rVert.
\label{eq47}
\end{equation}
Moreover, we can also apply the same procedure of normalizing the product after some steps and keep track of the normalization factor to calculate this product numerically. Finally, with the 
Shannon entropy rates (\ref{eq41}) and (\ref{eq47}) we obtain the rate of mutual information from (\ref{entZ2}) and (\ref{mutualdef}). 


In Fig. \ref{fig4}, we show the numerically obtained rate of mutual information for two sets of the kinetic parameters of the discrete time version of the one spin system of Fig. \ref{fig1} as a function of the time spacing $\tau$. 
For small $\tau$, the rate of mutual information 
shows a linear behavior, which we can extrapolate in order to obtain the continuous time rate of mutual information $\i$. The result has been shown in Fig. \ref{fig3}. A more
efficient numerical method to obtain $\i$, which generalizes the above discussion to the continuous time case, is introduced next.      

\subsection{Continuous time}
\label{5.2}

We consider the continuous time trajectory $\{Z(t)\}_{0}^{T}$ that stays in state $Z_n$ during the waiting time $\tau_n$. The number of jumps $N$ is a random functional of the trajectory and the time interval $T=\sum_{n=0}^{N}\tau_n$ is fixed.
The main difference, in relation to the discrete time case, is the presence of the exponentially distributed waiting times in the probability density of the continuous time trajectory, which is written as 
\begin{equation}
\mathcal{P}[\{Z(t)\}_{0}^{T}]=   \exp(-\lambda_{Z_N}\tau_N)\left[\prod_{n=1}^{N}w_{Z_{n-1}Z_{n}}\exp(-\lambda_{Z_{n-1}}\tau_{n-1})\right] P(Z_0). 
\end{equation}
where $P(Z_0)$ is the initial probability distribution. For $Z_n=(\alpha,i)$, the escape rate is
\begin{equation}
\lambda_{Z_n}\equiv \sum_{j\neq i}w_{ij}^\alpha+\sum_{\beta\neq \alpha}w_{i}^{\alpha\beta}.
\end{equation} 
Furthermore for  $Z_{n+1}=(\beta,j)$ the transition rates are $w_{Z_nZ_{n+1}}= w_{ij}^{\alpha\beta}$.

\begin{figure}
\centering
\includegraphics{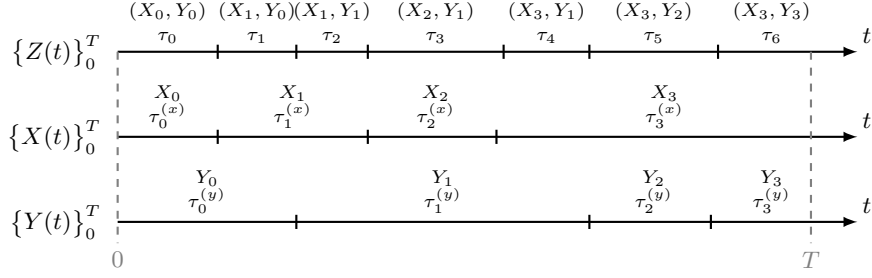}
\caption{Example of continuous time-series where the $Z$ process jumps  $6$ times and the $X$ and $Y$ process each jumps $3$ times, i.e., $N_x=N_y=3$.} 
\label{fig5}
\end{figure}

As illustrated in Fig. \ref{fig5}, the path $\{Z(t)\}_{0}^{T}$ has $N_x$ jumps for which the variable $X$ changes and $N_y$ jumps for which the variable $Y$ changes. Due to the bipartite form of the network of states, there are no jumps where
both variables change, which implies $N=N_x+N_y$. We denote the time intervals between jumps for the trajectory $\{X(t)\}_{0}^{T}$ by $\tau^{x}_n$, with $n=0,\ldots,N_x$. Similarly, for the trajectory
$\{Y(t)\}_{0}^{T}$ we have   $\tau^{y}_n$, with $n=0,\ldots,N_y$. In Fig. \ref{fig5}, an example of a trajectory with $N=6$ jumps is shown.  

The random matrix $\boldsymbol{\mathcal{T}}(Y_{n},Y_{n-1})$ is defined by its elements $\boldsymbol{\mathcal{T}}(Y_{n},Y_{n-1})_{X_n,X_{n-1}}$, which are the transition rate $w_{Z_{n-1}Z_{n}}$ if $Z_{n-1}\neq Z_{n}$ and 
$-\lambda_{Z_n}$ otherwise. More precisely, we can define $\boldsymbol{\mathcal{T}}(Y_{n},Y_{n-1})$ using its relation with the matrix $\boldsymbol{T}(Y_n,Y_{n-1})$, defined in (\ref{defTy}), which is
\begin{equation}
\boldsymbol{\mathcal{T}}(Y_n,Y_{n-1})\equiv \frac{1}{\tau}(\boldsymbol{T}(Y_n,Y_{n-1})-\mathbb{I}_x\delta_{Y_{n-1}Y_n}),
\end{equation}  
where $\mathbb{I}_x$ is the $\Omega_x\times\Omega_x$ identity matrix and $\delta_{Y_{n-1}Y_n}$ is the Kronecker delta. In addition, we define the matrix
\begin{equation}
\boldsymbol{\mathcal{F}}_{Y_n}(\tau)\equiv\exp\left(\boldsymbol{\mathcal{T}}(Y_{n},Y_{n})\tau\right).
\end{equation} 
Similarly to the discrete time case, for which equation (\ref{prodY}) holds, from the master equation, we obtain
\begin{align}
 \mathcal{P}[\{Y(t)\}_{0}^{T}]={} & \mathbf{V} \boldsymbol{\mathcal{F}}_{Y_{N_y}}(\tau_{N_y}^{(y)})\boldsymbol{\mathcal{T}}(Y_{N_y},Y_{N_y-1})\boldsymbol{\mathcal{F}}_{Y_{N_y-1}}(\tau_{N_y-1}^{(y)})\nonumber\\
& \times\ldots \boldsymbol{\mathcal{T}}(Y_2,Y_1)\boldsymbol{\mathcal{F}}_{Y_1}(\tau_1^{(y)})\boldsymbol{\mathcal{T}}(Y_1,Y_0)\boldsymbol{\mathcal{F}}_{Y_0}(\tau_0^{(y)}) \mathbf{P}_{Y_0}.
\label{prodcontY}
\end{align}

Moreover, the same expression is valid for the probability density of the $X$ time-series, i.e., 
\begin{align}
 \mathcal{P}[\{X(t)\}_{0}^{T}]={} & \mathbf{V} \boldsymbol{\mathcal{F}}_{X_{N_x}}(\tau_{N_x}^{(x)})\boldsymbol{\mathcal{T}}(X_{N_x},X_{N_x-1})\boldsymbol{\mathcal{F}}_{X_{N_x-1}}(\tau_{N_x-1}^{(x)})\nonumber\\
 &\times\ldots \boldsymbol{\mathcal{T}}(X_2,X_1)\boldsymbol{\mathcal{F}}_{X_1}(\tau_1^{(x)})\boldsymbol{\mathcal{T}}(X_1,X_0)\boldsymbol{\mathcal{F}}_{X_0}(\tau_0^{(x)}) \mathbf{P}_{X_0}.
\label{prodcontX}
\end{align}
The matrix $\boldsymbol{\mathcal{T}}(X_{n},X_{n-1})_{Y_n,Y_{n-1}}$ is now defined as  
\begin{equation}
\boldsymbol{\mathcal{T}}(X_n,X_{n-1})\equiv \frac{1}{\tau}(\boldsymbol{T}(X_n,X_{n-1})-\mathbb{I}_y\delta_{X_{n-1}X_n}),
\end{equation} 
where $\boldsymbol{T}(X_n,X_{n-1})$ is given by (\ref{defTx}) and $\mathbb{I}_y$ is the $\Omega_y\times \Omega_y$ identity matrix. The matrix $\boldsymbol{\mathcal{F}}_{X_n}(\tau)$ is defined as 
\begin{equation}
\boldsymbol{\mathcal{F}}_{X_n}(\tau)\equiv\exp\left(\boldsymbol{\mathcal{T}}(X_{n},X_{n})\tau\right).
\end{equation} 

In order to calculate the Shannon entropy rates a procedure similar to the discrete time case method can be used: we generate a long continuous time series, with the waiting times, $\{Z^*(t)\}_{0}^{T}$, with $N^*=N^*_x+N^*_y$ jumps, 
and estimate the non-Markovian Shannon entropy rates through the expressions 
\begin{align}
& H_Y\simeq -\frac{1}{T} \ln \left\lVert \boldsymbol{\mathcal{F}}_{Y^*_{N_y^*}}(\tau_{N_y^*}^{(y)}) \prod_{n=1}^{N_y^*}\boldsymbol{\mathcal{T}}(Y^{*}_n,Y^{*}_{n-1})\boldsymbol{\mathcal{F}}_{Y^*_{n-1}}(\tau_{n-1}^{(y)})\right\rVert,\nonumber\\
& H_X\simeq -\frac{1}{T} \ln \left\lVert \boldsymbol{\mathcal{F}}_{X^*_{N_x^*}}(\tau_{N_x^*}^{(x)}) \prod_{n=1}^{N_x^*}\boldsymbol{\mathcal{T}}(X^{*}_n,X^{*}_{n-1})\boldsymbol{\mathcal{F}}_{X^*_{n-1}}(\tau_{n-1}^{(x)})\right\rVert.
\end{align}
We are assuming that $N_x^*$ and $N_y^*$ are large, so that boundary terms can be neglected and we can use any matrix norm. These products are also numerically calculated by normalizing after a certain number of steps and 
keeping track of the normalization factors. The result obtained with the continuous time method for the one spin system of Fig. \ref{fig1} can be seen in Fig. \ref{fig3}. This method is more direct because for discrete time 
we have to obtain the result as a function of $\tau$ and then extrapolate for $\tau\to 0$. Moreover, when the probabilities of not jumping in discrete time are large, the continuous time
method is computationally cheaper. 

The continuous time method we presented above is not restricted to the bipartite networks we consider in this paper: it could be applied for other kinds of coarse-graining. The method only depends on the 
fact that we can write the probability density of a trajectory as a product of random matrices. 
\section{Four-state system}
\label{sec6}

\begin{figure}[h]
\centering
\includegraphics{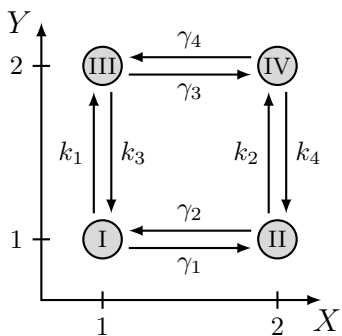}
\caption{General four-state model.} 
\label{fig6}
\end{figure}

We now illustrate the main results of this paper, namely, the analytical upper bound and the continuous time numerical method, by considering the  general four-state network shown in Fig. \ref{fig6}, for which  
the one spin system of Fig. \ref{fig1} is a particular example. Since $\Omega_x=\Omega_y=2$, there are four $\boldsymbol{\mathcal{T}}(Y_{n},Y_{n-1})$ and four $\boldsymbol{\mathcal{T}}(X_{n},X_{n-1})$ matrices, each of which is a two by two matrix.
For the sake of clarity, let us write these matrices explicitly. Using the superscript $(y)$ for the $\boldsymbol{\mathcal{T}}(Y_{n},Y_{n-1})$ matrices and $(x)$ for the $\boldsymbol{\mathcal{T}}(X_{n},X_{n-1})$ matrices, they are given by:
\begin{equation}
\boldsymbol{\mathcal{T}}^{(y)}(1,1)=\begin{pmatrix}
  			
  	-\gamma_1-k_1	& \gamma_2\\
	\gamma_1	& -\gamma_2-k_2 
		 
\end{pmatrix},\qquad
\boldsymbol{\mathcal{T}}^{(y)}(1,2)=\begin{pmatrix}
  			
  	k_3	& 0\\
	0	& k_4 
		 
\end{pmatrix},
\end{equation} 
\begin{equation}
\boldsymbol{\mathcal{T}}^{(y)}(2,1)=\begin{pmatrix}
  			
  	k_1	& 0\\
	0	& k_2 
		 
\end{pmatrix},\qquad
\boldsymbol{\mathcal{T}}^{(y)}(2,2)=\begin{pmatrix}
  			
  	-\gamma_3-k_3	& \gamma_4\\
	\gamma_3	& -\gamma_4-k_4 
		 
\end{pmatrix},
\end{equation} 
\begin{equation}
\boldsymbol{\mathcal{T}}^{(x)}(1,1)=\begin{pmatrix}
  			
  	-\gamma_1-k_1	& k_3\\
	k_1	& -\gamma_3-k_3 
		 
\end{pmatrix},\qquad
\boldsymbol{\mathcal{T}}^{(x)}(1,2)=\begin{pmatrix}
  			
  	\gamma_2	& 0\\
	0	& \gamma_4 
		 
\end{pmatrix},
\end{equation} 
\begin{equation}
\boldsymbol{\mathcal{T}}^{(x)}(2,1)=\begin{pmatrix}
  			
  	\gamma_1	& 0\\
	0	& \gamma_3 
		 
\end{pmatrix},\qquad
\boldsymbol{\mathcal{T}}^{(x)}(2,2)=\begin{pmatrix}
  			
  	-\gamma_2-k_2	& k_4\\
	k_2	& -\gamma_4-k_4 
		 
\end{pmatrix}.
\end{equation} 
In the following we treat two simple cases for which the rate of mutual information acquires a simple form in some limit.

\subsection{$Y$ following $X$}

Here we consider $k_1=k_4=0$. For this choice of rates a jump in the $Y$ process can happen only after a jump in the $X$ process. 
In this sense, $Y$ follows $X$. Calculating the stationary probability distribution, we obtain for the upper bound on the rate of mutual
information (\ref{ubi}) the expression 
\begin{equation}
\i^{(u)}= \frac{k_2 k_3\gamma}{2[(k_3+k_2)\gamma+k_2k_3]}\left(\ln\frac{k_2+2\gamma}{\gamma}+\ln\frac{k_3+2\gamma}{\gamma}\right),
\label{upperbound1}
\end{equation}   
where $\gamma_1= \gamma_2=\gamma_3= \gamma_4=\gamma$. If we further assume $k_2=k_3=k$ and $k\gg\gamma$, the rate of mutual information can be obtained with the following heuristic argument. A typical time series of the full process is an alternating sequence of long time intervals of size 
$1/\gamma$ with short time intervals  of size $1/k$. If we know the $X$ time series, we can predict in which of the $k/\gamma$ intervals of size $1/k$ the $Y$ jumps will take place. Since this information
amounting to $\ln k/\gamma$ occurs at the rate $\gamma$ of the $X$ jumps, we obtain that for $k\gg\gamma$
\begin{equation}
\i\simeq \gamma\ln \frac{k}{\gamma}.
\end{equation}         
More generally, for $k_1\neq k_3$, from the same kind of argument, we obtain
\begin{equation}
\i\simeq \frac{\gamma}{2}\left( \ln \frac{k_2}{\gamma}+\ln \frac{k_3}{\gamma}\right).
\end{equation}
This expression is in agreement with the upper bound (\ref{upperbound1}) in the limit $k_2,k_3\gg\gamma$.

\begin{figure}[h]
\centering	
\includegraphics{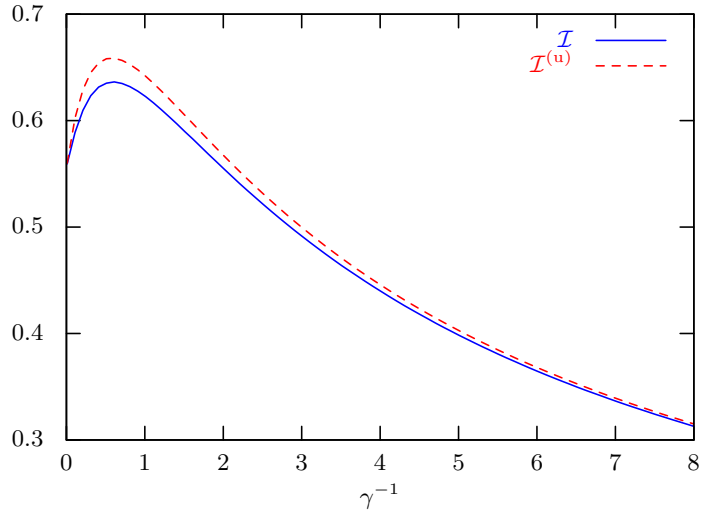}
\caption{Numerically obtained rate of mutual information $\i$ compared to the upper bound $\i^{(u)}$ (\ref{upperbound1}) as a function of $\gamma^{-1}$ for
$\gamma_1=\gamma_2=\gamma_3=\gamma_4=\gamma$, $k_1=k_4=0$, $k_3= 4$, and $k_2=1$. 
} 
\label{fig7}
\end{figure}

Moreover, we can also understand the rate of mutual information in the limit $\gamma\gg k_2,k_3$. This corresponds to the case where the $X$ process becomes Markovian and much faster than the $Y$ process, therefore, as discussed in Sect. \ref{sec4}
the rate of mutual information should saturate the upper bound. Suppose that we know the $Y$ time series. In the time interval between two $Y$ jumps there are many $X$ jumps and we have no information about the $X$ state
during this time interval. When a $Y$ jump takes place, we know the state $X$ with absolute precision, i.e., if the $Y$ jump is $1\to2$ ($2\to1$)  then the $X$ state is $2$ ($1$). Furthermore, since the $X$ jumps are fast compared to
$k_2,k_3$, the time interval between two $Y$ jumps is long enough for the $X$ process to decorrelate, so that the information obtained with an $Y$ jump is completely new. The complete knowledge of a binary random variable accounts for $\ln 2$ 
of mutual information. The average rate of $Y$ transitions is given by $k_3P_{III}+k_2P_{II}= k_2k_3/(k_2+k_3)$, where $P_{II}$ and $P_{III}$ denote the stationary probabilities of the states $II$ and $III$ defined in Fig. \ref{fig6}. This leads
to the expression 
\begin{equation}
\i \simeq \frac{k_2k_3}{k_2+k_3}\ln2,
\label{mutuheu1}
\end{equation}
valid for $\gamma\gg k_2,k_3$. As expected, this form is also in agreement 
with the upper bound (\ref{upperbound1}) in the respective limit. Fig. \ref{fig7}, where we compare the analytical upper bound with the numerical result, demonstrates that in the limits $k_2k_3\gg \gamma$ and
$\gamma\gg k_2,k_3$ the upper bound and the numerical result indeed tend to the same value.

\subsection{Equilibrium model}

\begin{figure}[h]
\centering
\includegraphics{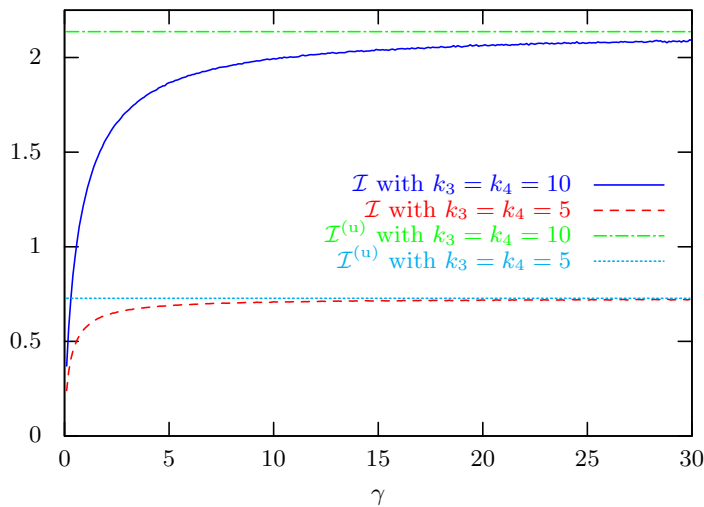}
\caption{Numerically obtained rate of mutual information $\i$ compared to the upper bound $\i^{(u)}$ (\ref{upperbound2}) as a function of $\gamma_1=\gamma_2=\gamma_3=\gamma_4=\gamma$.
The other parameters are $k_3= k_1= 1$ and $k_4=k_2$, thus enforcing equilibrium.}
\label{fig8}
\end{figure}

As a second example, we consider a network in equilibrium for which the rate of mutual information is nevertheless non-zero. In Fig. \ref{fig6},  we set $\gamma_1= \gamma_2=\gamma_3= \gamma_4= \gamma$, $k_3= k_1$, and $k_4= k_2$. 
For this choice of rates detailed balance is fulfilled because the product of the transition rates for the clockwise cycle equals the product of the transition rates for the counterclockwise cycle. Moreover, in the stationary 
state all states are equally probable. The upper bound on the rate of mutual information (\ref{ubi}) is independent of $\gamma$ and given by  
\begin{equation}
\i^{(u)}= \frac{1}{2}(k_1+k_2)\left(\ln 2- H(\epsilon)\right),
\label{upperbound2}
\end{equation}
where $\epsilon\equiv k_1/(k_1+k_2)$ and $H(\epsilon)\equiv-\epsilon\ln \epsilon -(1-\epsilon)\ln (1-\epsilon)$. As we show in Fig. \ref{fig8}, the rate of mutual information tends to the upper bound in the limit $\gamma\gg k_1,k_2$. This is again
in agreement with the discussion at the end of Sect. \ref{sec4}, since  the $X$ process is Markovian and, in the limit $\gamma\gg k_1,k_2$, much faster than the $Y$ process. Moreover, similarly to the way we obtained the result (\ref{mutuheu1})
for the previous model, the rate of mutual information can be easily explained in this limit. The difference in relation to the previous explanation is that when an $Y$ jump occurs the mutual information about the $X$ state is $\ln 2-H(\epsilon)$. 
This happens because if a $Y$ jump occurs, then the probability of $X$ being in state $1$ is $\epsilon$ and in state $2$ is $1-\epsilon$. As the average rate of a $Y$ jump is simply $(k_1+k_2)/2$, we obtain
\begin{equation}
\i\simeq \frac{1}{2}(k_1+k_2)\left(\ln 2- H(\epsilon)\right),
\label{mutuheu2}
\end{equation}
which is equal to the upper bound (\ref{upperbound2}).  
    
More generally, if the only restrictions are $\gamma_1= \gamma_2=\gamma_3= \gamma_4= \gamma$ and $\gamma\gg k_1,k_2,k_3,k_4$, then from the same kind of argument we obtain
\begin{align}
I & \simeq (P_Ik_1+P_{II}k_2)(\ln 2-H(\epsilon_1))+(P_{III}k_3+P_{IV}k_4)(\ln 2-H(\epsilon_2))\nonumber\\
  & = \frac{(k_1+k_2)(k_3+k_4)}{2(k_1+k_2+k_3+k_4)}(2\ln 2-H(\epsilon_1)-H(\epsilon_2)),
\end{align}
where $\epsilon_1= k_1/(k_1+k_2)$ and $\epsilon_2= k_3/(k_3+k_4)$. This more general expression accounts for the results (\ref{mutuheu1}) and (\ref{mutuheu2}).

\section{Summary}
\label{sec7}

In this paper we have addressed the problem of calculating the rate of mutual information between two 
coarse-grained processes that together fully specify a continuous time Markov process. To this end,
we have developed a numerical method to estimate the Shannon entropy rate of hidden Markov processes 
from a continuous time series, generalizing the numerical method used in the discrete time case
\cite{holl06,jacq08,rold12}. Moreover, for the class of bipartite Markov processes we considered in 
this paper, we have obtained an expression for an upper bound on the rate of mutual information in 
terms of the stationary probability distribution. While this expression has some formal
similarity with the one for the rate of thermodynamic entropy production, it has become clear that 
these two rates, in general,  are not related through a simple inequality.

As applications of the theory developed here we have studied three four-state systems each of which
can serve as illustrating, {\sl inter alia},  the apparent independence of the rate
of mutual information from the rate of thermodynamic entropy production. First,
the one spin system with time-varying magnetic field is arguably the simplest case which shows that
in an non-equilibrium steady state the rate of mutual information is not bounded by the
dissipation rate. Second, for a four state network for which some transition rates are zero,
the rate of mutual information is still well defined whereas the thermodynamic entropy production
is not since the latter requires that each backward transition is possible with a finite rate as well.
Third, a four state system in equilibrium with zero thermodynamic entropy production can still
have non-zero rate of mutual information. Moreover, in these four-state systems it is typically  
possible to find, and to understand in simple terms, a limiting case for the rates such that the 
analytical upper bound on the rate of mutual information becomes saturated.

On the mathematical side, finding a general expression for the rate of mutual information at least for 
the bipartite case on which we focused is most likely as hard a problem as finding one for the
Shannon entropy rate of a non-Markovian process. For interesting physical perspectives, the rate
of mutual information could become particularly relevant for the emerging theories of both autonomous
information machines and cellular sensing systems. In both cases, one could suspect that even though
there is no simple bound between the information-theoretic and the thermodynamic rate of entropy
production in general, in more specific settings these two quantities might obey relations still 
to be uncovered. The algorithm described here to calculate the former will help in generating the 
necessary data for any specific model network efficiently.

\begin{acknowledgements}
Support by the ESF through the network EPSD  is gratefully acknowledged.
\end{acknowledgements}

\appendix

\section{Detailed derivation of the analytical upper bound}

The first upper bound $H(Y_2|Y_1)$ can be easily calculated by using the conditional probability 
\begin{equation}
P(Y_2|Y_1)= \frac{\sum_{X_1}P(Y_2,Y_1,X_1)}{P(Y_1)}= \frac{\sum_\alpha P_i^{\alpha}w^{\alpha}_{ij}\tau}{P_i},
\end{equation}
where $Y_2\neq Y_1$. We here performed the substitutions $X_1\to\alpha$, $Y_1\to i$, and $Y_2\to j$. Using this formula in (\ref{conddef}) we obtain 
\begin{eqnarray}
H(Y_2|Y_1)=-\sum_{i,\alpha}P_i^\alpha\sum_{j\neq i}w_{ij}^\alpha\left(\ln \tau+\ln\frac{\sum_\beta P_i^\beta w_{ij}^\beta}{P_i}-1\right)+\textrm{O}(\tau).
\label{hy1}
\end{eqnarray} 

Moreover, $H(Y_{N+1}|Y_{N},\ldots,Y_1)$ up to order $\tau$ is given by the above formula for any finite $N$. In order to demonstrate this we first rewrite (\ref{conddef}) as
\begin{align}
& H(Y_{N+1}|Y_{N},\ldots,Y_1) \nonumber\\
& = -\frac{1}{\tau}\sum_{Y_{N+1}\neq Y_N}\sum_{Y_N\ldots Y_1}P(Y_{N+1},Y_N,\ldots,Y_1)\ln P(Y_{N+1}|Y_N,\ldots,Y_1)\nonumber\\
& -\frac{1}{\tau}\sum_{Y_N\ldots Y_1}P(Y_{N},Y_N,\ldots,Y_1)\ln P(Y_{N}|Y_N,\ldots,Y_1),
\end{align}
where $P(Y_N,Y_N,\ldots,Y_1)$ denotes the probability of having a sequence for which $Y_{N+1}=Y_N$. For $Y_{N+1}\neq Y_N$, the expression of the conditional probability $P(Y_{N+1}|Y_N,\ldots,Y_1)$
has at least one transition probability term of order $\tau$. Therefore, as $P(Y_{N+1}|Y_N,\ldots,Y_1)$ is at least a term of order $\tau$, it is convenient to further rewrite the above expression as  
\begin{align}
& H(Y_{N+1}|Y_{N},\ldots,Y_1) =-\frac{1}{\tau}\sum_{Y_{N+1}\neq Y_N}\sum_{Y_N}P(Y_{N+1},Y_N)\ln \tau \nonumber\\
& -\frac{1}{\tau}\sum_{Y_{N+1}\neq Y_N}\sum_{Y_N\ldots Y_1}P(Y_{N+1},Y_N,\ldots,Y_1)\ln \frac{P(Y_{N+1}|Y_N,\ldots,Y_1)}{\tau}\nonumber\\
& -\frac{1}{\tau}\sum_{Y_N\ldots Y_1}P(Y_{N},Y_N,\ldots,Y_1)\ln P(Y_{N}|Y_N,\ldots,Y_1), 
\label{ure}
\end{align}
where in the first line we summed over the variables $Y_1,\ldots,Y_{N-1}$. The three following relations are important for the subsequent derivation. 
First, for $Y_{N+1}\neq Y_N$,
\begin{equation}
P(Y_{N+1}, Y_N,\ldots,Y_1) = \left\{
  \begin{array}{l l}
    P(Y_{N+1}, Y_N) + \textrm{O}(\tau^2) & \quad \text{if $Y_N= Y_{N-1}=\ldots=Y_1$}\\
    \textrm{O}(\tau^2) & \quad \text{otherwise}.
  \end{array} \right.
\end{equation}
Moreover, 
\begin{equation}
P(Y_N,\ldots,Y_1) = \left\{
  \begin{array}{l l}
    P(Y_N) + \textrm{O}(\tau) & \quad \text{if $Y_N= Y_{N-1}=\ldots=Y_1$}\\
    A\tau^\eta+ \textrm{O}(\tau^{\eta+1}) & \quad \text{otherwise},
  \end{array} \right.
\end{equation}
where $\eta\ge1$ is an integer and $A$ is a constant independent of $\tau$. Finally, the conditional probability distribution fulfills  
\begin{equation}
P(Y_{N+1}| Y_N,\ldots,Y_1) = \left\{
  \begin{array}{l l}
    P(Y_{N+1}|Y_N) + \textrm{O}(\tau^2) & \quad \text{if $Y_N= Y_{N-1}=\ldots=Y_1$}\\
    B\tau^\nu+ \textrm{O}(\tau^{\nu+1})& \quad \text{otherwise},
  \end{array} \right.
\end{equation}
where $\nu\ge1$ is an integer and $B$ is a constant independent of $\tau$. With these three relations, the term in the second line in equation (\ref{ure}) becomes
\begin{align}
& \frac{1}{\tau}\sum_{Y_{N+1}\neq Y_N}\sum_{Y_N\ldots Y_1}P(Y_{N+1},Y_N,\ldots,Y_1)\ln \frac{P(Y_{N+1}|Y_N,\ldots,Y_1)}{\tau}\nonumber\\
& = \frac{1}{\tau}\sum_{Y_{N+1}\neq Y_N}\sum_{Y_N} P(Y_{N+1},Y_N)\ln \frac{P(Y_{N+1}|Y_N)}{\tau}+\textrm{O}(\tau),
\label{ure1}
\end{align}
where we used $\tau^{\nu+\eta-1}\ln \tau^{\nu-1}\in\textrm{O}(\tau)$. For the term in the third line in equation (\ref{ure}) we need the relations,
\begin{equation}
P(Y_{N}|Y_N,\ldots,Y_1)= 1-\sum_{Y_{N+1}\neq Y_{N}} P(Y_{N+1}|Y_N,\ldots,Y_1)
\end{equation}
and 
\begin{equation}
P(Y_{N},Y_N,\ldots,Y_1)= P(Y_N,\ldots,Y_1) \left(1-\sum_{Y_{N+1}\neq Y_{N}} P(Y_{N+1}|Y_N,\ldots,Y_1)\right)
\end{equation}
which lead to
\begin{equation}
\frac{1}{\tau}\sum_{Y_N\ldots Y_1}P(Y_{N},Y_N,\ldots,Y_1)\ln P(Y_{N}|Y_N,\ldots,Y_1)= \frac{1}{\tau}\sum_{Y_{N+1}\neq Y_N} P(Y_{N+1},Y_{N})+\textrm{O}(\tau).
\label{ure2}
\end{equation}
Inserting  (\ref{ure1}) and (\ref{ure2}) in (\ref{ure}) we obtain
\begin{equation}
H(Y_{N+1}|Y_{N},\ldots,Y_1)= H(Y_{N+1}|Y_{N})+ \textrm{O}(\tau).
\end{equation}
Therefore, since the $Y$ process is stationary, from (\ref{hy1}), we obtain for any finite $N$ 
\begin{eqnarray}
H(Y_{N+1}|Y_{N},\ldots,Y_1)=-\sum_{i,\alpha}P_i^\alpha\sum_{j\neq i}w_{ij}^\alpha\left(\ln \tau+\ln\frac{\sum_\beta P_i^\beta w_{ij}^\beta}{P_i}-1\right)+\textrm{O}(\tau).
\end{eqnarray} 
Applying the same method to the $X$ process we get,
\begin{eqnarray}
H(X_{N+1}|X_{N},\ldots,X_1)=-\sum_{i,\alpha}P_i^\alpha\sum_{\beta\neq \alpha}w_{i}^{\alpha\beta}\left(\ln \tau+\ln\frac{\sum_j P_j^\alpha w_{j}^{\alpha\beta}}{P^\alpha}-1\right)+\textrm{O}(\tau).
\end{eqnarray}


\end{document}